\documentstyle[12pt]{article}

\begin{document}

\title{SCALAR FIELD COSMOLOGIES WITH VISCOUS FLUID}

\author{Luis P. Chimento and Alejandro S. Jakubi      \\
{\it Departamento de F\'{\i}sica,  }\\
{\it  Facultad de Ciencias Exactas y Naturales, }\\
{\it Universidad de Buenos Aires }\\
{\it  Ciudad  Universitaria,  Pabell\'{o}n  I, }\\
{\it 1428 Buenos Aires, Argentina.}}

\maketitle

\begin{abstract}

We investigate cosmological models with a free scalar field and
a viscous fluid. We find exact solutions for a linear and nonlinear
viscosity pressure. Both yield singular and bouncing solutions. In the first
regime, a de Sitter stage is asymptotically stable, while in the second case we
find power-law evolutions for vanishing cosmological constant.

\end{abstract}

\vskip 3cm


\section{Introduction}

Studies of the dynamics of a scalar field together with other forms of
matter usually assume that this behaves as a perfect fluid. Few authors have
taken into account dissipative effects  \cite{Sin88} \cite{Sin90} \cite{Roy92}
\cite{Roy93}, even though processes like particle creation might have been
important in the early universe.
 In this paper we study the evolution of a universe filled with
a massless minimally coupled scalar field and a viscous fluid. We find exact
solutions of the Einstein equations in a Robertson-Walker metric and we
analyse their asymptotic stability by means of the Lyapunov method
\cite{Kra}.

\section{The model}

We investigate the evolution of a universe filed with a scalar field
and a viscous fluid. The scalar field $\phi $ is free and minimally
coupled to gravity, so that it obeys the equation $\Box \phi =0$.
In the case of the homogeneous isotropic Robertson-Walker
metric

\begin{equation}\label{1}
ds^2=dt^2-a^2(t)\left [{\frac{dr^2}{1-kr^2}}+r^2(d\theta ^2+\sin
{}^2\theta d\phi ^2)\right]
\end{equation}

\noindent
the scalar field equation becomes

\begin{equation}
\label{100}
\ddot \phi +3H\dot \phi =0
\end{equation}

Besides, only the bulk viscosity needs to be considered. Thus we replace in
the Einstein equations the equilibrium pressure $p$ by an effective pressure
\cite{Wei71}

\begin{equation}\label{2}
3H^2=\frac {1}{2}\dot\phi^2+\rho -3{\frac k{a^2}}+\Lambda
\end{equation}
\begin{equation} \label{3}
2\dot H+ 3H^{2} = -\frac{1}{2}\dot\phi^2 - p -\sigma-\frac{k}{a^2}+\Lambda
\end{equation}

\noindent where $H=\dot a/a$, $^{\cdot }=d/dt$, $\rho $ is the energy
density, $\sigma $ is the viscous pressure, $\Lambda$ is the cosmological
constant and we use units $c=8\pi G=1$.
As equation of state we take

\begin{equation}
\label{4}
p = ( \gamma -1 ) \rho
\end{equation}

\noindent  with a constant adiabatic index $0\le\gamma \le 2$, and we assume
that $\sigma $ has the constitutive equation

\begin{equation}\label{45}
\sigma = - 3 \zeta H
\end{equation}

\noindent where $\zeta \ge 0$ is the bulk viscosity coefficient.
Thus, we must solve equation (\ref{100}) together with the Einstein equations
(\ref{2})(\ref{3}). Equation (\ref{100}) has the first integral

\begin{equation} \label{5}
\dot\phi=\frac{C}{a^3}\qquad
\end{equation}

\noindent
where $C$ is an arbitrary integration constant, and we are interested in the
case that $C\neq0$.
Then, adding (\ref{2}) and (\ref{3}), and eliminating $\rho$, we arrive at

\begin{equation} \label{6}
2\dot H+3\gamma H^2-3\zeta H=-\frac{2-\gamma}{2}\frac{C^2}{a^6}+
\left(2-3\gamma\right)\frac{k}{a^2}+\gamma\Lambda
\end{equation}

To find exact solutions of this equation we make the change of variable
$a=v^\nu$, and we consider two cases:  $\zeta$ constant and $\zeta \sim
H$.
We assume now that $\zeta$ is a constant. This is a good approximation when
the thermodynamical variables do not change too much. This has been studied in
several papers \cite{Tre} \cite{Hel73} \cite{Hel74} \cite{Sus}
\cite{Roy83}. Choosing $\nu=2/(3\gamma)$, the equation (\ref{6}) becomes

\begin{equation} \label{7}
\ddot v+\frac{3}{2} \zeta\dot v+M v^m+N v^n-\frac{\gamma\Lambda}{2\nu}v=0
\end{equation}

\noindent
where $M=(3/8)\gamma\left(2-\gamma\right)C^2$, $m=1-4/\gamma$,
$N=(3/4)\gamma(3\gamma-2)k$, $n=1-4/(3\gamma)$.
It is linear for $\gamma=2$ and $k=0$, so that we can find its general
solution. In this case, the evolution of the scale factor is the same as in a
model without the scalar field and a bulk viscosity coefficient $\zeta/2$
\cite{Tre} \cite{Hel73}. Thus, we just quote the solutions without further
analysis.

\bigskip
\noindent
Let us consider first that $\Lambda=0$. We find
\bigskip

\begin{equation} \label{8}
a(t)=\left[A \exp\left(\frac{3}{2}\zeta t\right)+B\right]^{1/3}
\end{equation}
\begin{equation} \label{9}
\Delta\phi(t)=\frac{2C}{3\zeta B}\ln\left[\frac{\exp\left(\frac{3}{2}\zeta
t\right)} {A\exp\left(\frac{3}{2}\zeta t\right)+B}\right]
\end{equation}
\begin{equation} \label{10}
\rho(t)=\frac{\frac{3}{2}\zeta^2 A^2\exp(3\zeta t)-C^2}
{2\left[A\exp\left(\frac{3}{2}\zeta t\right)+B\right]^2}
\end{equation}

\noindent
where $A$ and $B$ are arbitrary integration constants. The requirement
$\rho\ge0$ implies that there is a minimum time for the validity of this
solution. Thus it describes a geodesically incomplete manifold, which in this
sense is singular \cite{Ger}.
Clearly, in its asymptotically de Sitter regime it is physically
justified the approximation of constant $\zeta$. However, for shorter times,
this assumption may not be justified.

The asymptotically de Sitter stage occurs for any other value of $\gamma$ and
$k$. In effect, assuming $a\sim\exp(H_0 t)$ for $t\to\infty$, with $H_0$ a
constant, we find that $H_0=\zeta/\gamma$. To verify the asymptotic stability
of this solution we use $a$ as the independent variable and we take the
Liapunov function $L=(H-H_0)^2$. Then, we find that

\begin{equation} \label{11}
L'=-3\gamma\frac{L}{a}+O\left(\frac{1}{a^3}\right)
\end{equation}

\noindent
is negative for large times.

Let us consider now that $\Lambda\neq0$.
There is a critical value for the cosmological constant
$\Lambda_0=-3\zeta^2/16$. Thus, we must distinguish several cases:

\noindent
Two kinds of solutions appear for $\Lambda>\Lambda_0$. Singular evolutions

\begin{equation} \label{12}
a(t)=a_0 e^{3\Delta t/4}\left[\sinh\left(\omega \Delta t
\right)\right]^{1/3}
\end{equation}

\noindent
or nonsingular evolutions

\begin{equation} \label{13}
a(t)=a_0 e^{3\Delta t/4}\left[\cosh\left(\omega \Delta t
\right)\right]^{1/3}
\end{equation}

\noindent
where $\omega=\left[3\left(\Lambda-\Lambda_0\right)\right]^{1/2}$, $\Delta
t=t-t_0$ and $a_0$, $t_0$ are integration constants. For the
scalar field we find

\begin{equation} \label{14}
\Delta
\phi(t)=-\frac{C}{B\mu}e^{(\omega-3\zeta/4)\Delta t}
{}_2F_1\left(1,\frac{1}{2}-\frac{3\zeta}{8\omega},\frac{3}{2}
-\frac{3\zeta}{8\omega},-\frac{A}{B}e^{2\omega \Delta t}\right)
\end{equation}

While $\rho\to (3/8)\zeta^2+(3/2)\zeta\omega>0$ for $t\to\infty$, in the
nonsingular solutions $\rho$ becomes negative before a minimum time. On the
other hand, for singular solutions, we find

\begin{equation} \label{15}
\rho\sim\left(\frac{1}{3}-\frac{C^2}{2a_0^3\omega^2}\right)\frac{1}{\Delta t^2}
\qquad \Delta t\to 0
\end{equation}

\noindent
so that the energy density may remain positive along all the evolution.

\noindent
In the case $\Lambda=\Lambda_0$, the evolution is always singular

\begin{equation} \label{16}
a(t)=a_0 (\Delta t)^{1/3} e^{3\Delta t/4}
\end{equation}

\noindent
In the case $\Lambda\le\Lambda_0$, the scale factor recollapses to a second
singularity

\begin{equation} \label{17}
a(t)=a_0 e^{3\Delta t/4}\left[\sin\left(|\omega| \Delta t\right)\right]^{1/3}
\end{equation}

\noindent
All singular evolutions have particle horizons because $a\sim \Delta t^{1/3}$
as $\Delta t\to 0$.

For other values of $\gamma$ and $k$, there is also a critical value of the
cosmological constant $\Lambda_0=-3\zeta^2/(4\gamma^2)$, so that for
$\Lambda>\Lambda_0$ an asymptotically de Sitter evolution occurs for two
values of $H$

\begin{equation} \label{18}
H_\pm=\frac{1}{2\gamma}\left[\zeta\pm\left(\zeta^2+\frac{4}{3}\gamma^2\Lambda
\right)^{1/2}\right]
\end{equation}

\noindent
We verify that the behavior $a\sim\exp(H_+t)$ is asymptotically stable by
means of the Liapunov function $L_+=(H-H_+)^2$, which satisfies $L_+'<0$ for
large times when $H>0$.

Nonlinear viscous effects has been shown to arise as a phenomenological
description of the effect of particle creation in the early universe
\cite{Ver}, and cosmological models with this kind of fluids has been
considered in \cite{Nov} and \cite{Rom}.

Here, we assume that $\zeta=\alpha H$, with a constant $\alpha$ such that
$\gamma>\alpha>0$, and we consider only expanding evolutions.
Following the same procedure as before, we arrive at the equation

\begin{equation} \label{20}
\ddot v+M v^m+N v^n-\frac{\gamma\Lambda}{2\nu}v=0
\end{equation}

\noindent
where now $\nu=2/[3(\gamma-\alpha)]$, $M=(3/8)(\gamma-\alpha)(2-\gamma)$,
$m=1-4/(\gamma-\alpha)$, $N=(3/4)(\gamma-\alpha)(3\gamma-2)k$ and
$n=1-4/[3(\gamma-\alpha)]$. As this case behaves as a conservative mechanical
system, we may obtain (at least formally) its general solution:

\begin{equation} \label{21}
\Delta t=\int\frac{dv}{\sqrt{2\left[E-V(v)\right]}}
\end{equation}

\noindent
where $E$ is an integration constant and

\begin{equation} \label{22}
V(v)=\frac{M}{m+1}v^{m+1}+\frac{N}{n+1}v^{n+1}-\frac{\gamma\Lambda}{4\nu}v^2
\qquad \gamma-\alpha\neq\frac{2}{3}
\end{equation}

\begin{equation} \label{23}
V(v)=\frac{M}{m+1}v^{m+1}+N\ln v-\frac{\gamma\Lambda}{4\nu}v^2
\qquad \gamma-\alpha=\frac{2}{3}
\end{equation}

A qualitative analysis of (\ref{21}) shows that expanding evolutions may begin
either at a singularity or a bounce. Both bounded and unbounded singular
solutions occur, but only for $\Lambda<0$ are there bouncing solutions that
reach a maximum.

We quote two cases for which the equation (\ref{20}) becomes linear, so that we
may obtain its general solution in closed form (for simplicity we take
$\Lambda=0$ ). The first one is $\gamma=2$ and $k=0$

\begin{equation} \label{24}
a(t)=\left(A\Delta t\right)^\nu
\end{equation}

\begin{equation} \label{25}
\Delta\phi(t)=\frac{C\Delta t^{1-3\nu}}{A^{3\nu}(1-3\nu)}
\end{equation}

\begin{equation} \label{26}
\rho(t)=\frac{3\nu^2}{\Delta t^2}-\frac{C^2}{2(A\Delta t)^{6\nu}}
\end{equation}

\noindent
The other solution is for $\gamma=2$, $\alpha=2/3$ and $k\neq0$. We find

\begin{equation} \label{27}
a(t)=\left|A\Delta t-2k\Delta t^2\right|^{1/2}
\end{equation}

\begin{equation} \label{28}
\Delta\phi(t)=-\frac{2C}{A^2}\frac{A-4k\Delta t}
{\left|A\Delta t-2k\Delta t^2\right|^{1/2}}
\end{equation}

\begin{equation} \label{29}
\rho(t)=\frac{3}{4}\frac{(A-4k\Delta t)^2}
{\left|A\Delta t-2k\Delta t^2\right|^2}-\frac{1}{2}\frac{C^2}
{\left|A\Delta t-2k\Delta t^2\right|^3}+\frac{3k}
{\left|A\Delta t-2k\Delta t^2\right|}
\end{equation}

We find also solutions of the form $a=a_0 \Delta t^\sigma$ for special values
of $\alpha$: a. $\gamma=2$, $\sigma=1$, $\alpha=(4/3)(1+k/a_0^2)$; b. $k=0$,
$\sigma=1/3$, $\alpha=\gamma+3(2-\alpha)C^2/(2a_0^2)-2$.

\section{Conclusions}

We have found exact solutions of the Einstein equations with a free scalar
field and a viscous fluid source, in a homogeneous isotropic metric. We have
considered cosmological models with two forms of the bulk viscosity pressure:
linear, with a constant coefficient, and cuadratic in the Hubble variable. For
the first case the picture is similar to the case without scalar field and we
show that the de Sitter evolution is asymptotically stable. Also, we find that
the physical requirement of positivity of the energy density of the fluid
makes some of these solutions geodesically incomplete.
For the case of a nonlinear viscous pressure we reduce the equations of this
model to that of a conservative mechanical system. Thus, we are able to give
its general solution in implicit form, and besides we show several cases for
which an explicit solution arise. Both, singular and bouncing solutions arise,
and we find cases in which the scale factor reach a maximum or grow without
bound. In the latter case, when $\Lambda=0$, the scale factor has a power
law behavior for large times.

Dissipative effects like particle production may have been important in the
early universe, for instance in the reheating period at the end of the
inflationary era. It is generally assumed that this stage of accelerated
evolution was driven by a self-interacting scalar field. Thus, we consider of
interest to investigate further the models of this paper, taking into account
also an interaction potential.

\end{document}